\documentclass[12pt]{article} 
\usepackage{graphicx} 
\def \b{{\cal B}} 
 
\def \bea{\begin{eqnarray}} 
\def \beq{\begin{equation}}

\def \eea{\end{eqnarray}} 
\def \eeq{\end{equation}}

\def \s{\sqrt{2}} 
 
\def\lsim{\mathrel{\rlap{\lower3pt\hbox{$\sim$}}\raise2pt\hbox{$<$}}}
\def\gsim{\mathrel{\rlap{\lower3pt\hbox{$\sim$}}\raise2pt\hbox{$>$}}}
\def\btopik{B \to \pi K}
\def\bd{B^0}
\def\ApNPqph{{\cal A}^{q} e^{i \Phi_q}}
\def\ApNPCqph{{\cal A}^{{ C}, q} e^{i \Phi_q^{C}}}
\def\ApNPCuph{{\cal A}^{{ C}, u} e^{i \Phi_u^{C}}}
\def\ApNPCdph{{\cal A}^{{ C}, d} e^{i \Phi_d^{C}}}
\def\ApNPuph{{\cal A}^{u} e^{i \Phi_u}}
\def\ApNPdph{{\cal A}^{d} e^{i \Phi_d}}
\def\ApNPcomb{{\cal A}^{comb} e^{i \Phi}}
\def\pewc{P_{ EW}^{C}}
\def\pew{P_{ EW}}
\def\pewcpnp{P_{ EW, NP}^{C}}
\def\pewpnp{P_{ EW, NP}}

\def\bra#1{\left\langle #1\right|}
\def\ket#1{\left| #1\right\rangle}

\begin{document} 
\begin{flushright}
TECHNION-PH-2009-08 \\ 
UdeM-GPP-TH-09-178 \\
EFI 09-15 \\ 
arXiv:0905.1495 [hep-ph] \\ 
May 2009 \\ 
\end{flushright} 
\centerline{\bf\boldmath DIAGNOSTIC FOR NEW PHYSICS IN $\btopik$ DECAYS
\footnote{To be submitted to Physics Letters B.}} 
\medskip
\centerline{Seungwon Baek}
\centerline{\it Institute of Basic Science and Department of Physics}
\centerline{\it Korea University, Seoul 136-701, Korea}
\medskip
\centerline{Cheng-Wei Chiang}
\centerline{\it Department of Physics and Center for Mathematics and
  Theoretical Physics}
\centerline{\it National Central University, Chungli, Taiwan 320,
  R.O.C. and}
\centerline{\it Institute of Physics, Academia Sinica, Taipei, Taiwan
  115, R.O.C.}
\medskip 
\centerline{Michael Gronau} 
\centerline{\it Physics Department, Technion, Haifa 32000, Israel.}
\medskip
\centerline{David London}
\centerline{\it Physique des Particules, Universit\'e de Montr\'eal}
\centerline{\it C. P. 6128, succ.\ centre-ville, Montr\'eal, QC, Canada
H3C 3J7}
\medskip 
\centerline{Jonathan L. Rosner} 
\centerline{\it Enrico Fermi Institute and Department of Physics,
  University of Chicago} 
\centerline{\it Chicago, IL 60637, U.S.A.} 
\bigskip 
\begin{quote} 
A recent analysis of $B \to \pi K$ decays concludes that
present data do not clearly indicate whether (i) the
standard model (or $\Delta I=0$ new physics) is sufficient,
or (ii) $\Delta I=1$ new physics is needed. We show that
these two possibilities can be distinguished by whether a sum
rule relating the CP asymmetries of the four $B \to \pi K$
decays is valid.  If case (i) is favored, the sum rule holds,
and one predicts $A_{CP}(\pi^0 K^0) = -0.15$, while in case
(ii) fits to new physics 
involving large values of a color-suppressed tree amplitude
entail $A_{CP}(\pi^0 K^0) = -0.03$.
The current experimental average $A_{CP}(\pi^0 K^0) = -0.01
\pm 0.10$ must be measured a factor of at least three times
more precisely in order to distinguish between the two cases.
\end{quote}
 
\leftline{\qquad PACS codes: 12.15.Hh, 12.15.Ji, 13.25.Hw,
14.40.Nd}
 
\newpage

Several suggestions have been made in the past few years that
data for $\btopik$ decays cannot be simply explained within
the Cabibbo-Kobayashi-Maskawa (CKM) framework.
In most cases
these claims suffered from uncertainties in QCD calculations
of hadronic matrix elements using a heavy-quark expansion.
The purpose of this note is to outline some of these claims,
proposing a way for using improved data in order to make a
more robust statement about new physics.

The four $\btopik$ decay amplitudes are related by
isospin \cite{Nir:1991cu},
\beq\label{Iso}
A(\bd \to \pi^- K^+) - \s A(B^+ \to \pi^0 K^+) + \s A(\bd \to
\pi^0 K^0) - A(B^+ \to \pi^+ K^0) = 0~.
\eeq
The four amplitudes may be written in terms of penguin
($P_{tc}$ and $P_{uc}$), tree ($T$), color-suppressed ($C$),
annihilation ($A$), color-favored electroweak penguin
($P_{EW}$), and color-suppressed electroweak penguin
($P_{EW}^C$) contributions \cite{Gronau:1994rj}:
\bea
\label{amp-+}
-A(B^0 \to \pi^- K^+) &=& P_{tc} + P_{uc} +
T +\frac23 \pewc ~, \\ 
\label{amp0+}
-\s A(B^+ \to \pi^0 K^+) &=& P_{tc} +
P_{uc} + T + C + A + \pew + \frac23 \pewc~, \\
\s A(B^0 \to \pi^0 K^0) &=& P_{tc}  + P_{uc}
- C - \pew - \frac13 \pewc~, \\
A(B^+ \to \pi^+ K^0) &=& P_{tc} + P_{uc} + A
-\frac13 \pewc ~.  
\eea
The two amplitudes $P_{tc}$ and $P_{uc}$ behave like isoscalars ($\Delta I=0$)
while the remaining terms are mixtures of isoscalar and isovector
($\Delta I=1$).  The terms $P_{tc}, \pew$ and $\pewc$ involve a CKM factor
$V^*_{tb}V_{ts}$ with weak phase $\pi$, while $P_{uc}, T$,
$C$ and $A$ contain $V^*_{ub}V_{us}$ with phase
$\gamma$. Each of these amplitudes has its own unknown strong
phase, but some strong phases can be related to each other
approximately.  Using flavor SU(3) symmetry, the amplitudes
$\pew$ and $\pewc$ are given to a good approximation in terms
of $T$ and $C$ \cite{EWPs}:
\bea
\pew & = & {3\over 4} {c_9 + c_{10} \over c_1 + c_2} R (T +
C) \!+\!  {3\over 4} {c_9 - c_{10} \over c_1 - c_2} R (T - C)
\approx {3\over 2} {c_9 + c_{10} \over c_1 + c_2} RT
~, \\
\label{pewc}
\pewc & = & {3\over 4} {c_9 + c_{10} \over c_1 + c_2} R (T +
C) \!-\!  {3\over 4} {c_9 - c_{10} \over c_1 - c_2} R (T - C)
\approx {3\over 2} {c_9 + c_{10} \over c_1 + c_2} RC
~.
\eea
Here, $c_i$ are Wilson coefficients \cite{BuraseffH} obeying
$(c_9 + c_{10})/(c_1 + c_2)\approx (c_9 - c_{10})/(c_1 -
c_2)$, while $R \equiv \left\vert (V_{tb}^* V_{ts})/(V_{ub}^*
V_{us})\right\vert= 48.9\pm 1.6$ \cite{Charles:2004jd}.  The
proportionality coefficient relating $\pew$ to $T$ and
$\pewc$ to $C$ is numerically of order one and negative:
$\delta_{EW}\equiv (3/2)[(c_9 + c_{10})/(c_1 + c_2)]R=
-0.60\pm 0.02$. SU(3)
breaking introduces a theoretical error of about $10\%$ in
the magnitude of this coefficient and an error of $5^\circ$
in its strong phase \cite{Neubert:1998re}.

The decays $\btopik$ provide nine measurements: four
branching fractions $\b$, four direct CP asymmetries $A_{
CP}$, and the mixing-induced CP asymmetry $S_{ CP}$ in
$\bd\to \pi^0K^0$.  Assuming values for the weak phases
$\beta$ and $\gamma$ as determined in a global CKM fit
\cite{Charles:2004jd}, these observables can be expressed in
terms of nine parameters: the magnitude of the five
independent amplitudes, $P_{tc}$, $P_{uc}$, $T$, $C$, $A$,
and their four relative strong phases. Neglecting the
amplitude $A$, which vanishes to leading order in $1/m_b$ and
$\alpha_s$ \cite{Bauer:2004tj}, one may perform a best fit
with two degrees of freedom.  Such a fit was made about two
years ago \cite{Baek:2007yy} with data available in early
2007. The fit neglected $P_{uc}$ but kept $\gamma$ as a free
parameter which was extracted successfully in agreement with
its determination in the CKM fit. A good fit, corresponding
to $\chi^2_{min}/d.o.f.=1/3~(80\%$ c.l.), required $|C/T| =
1.6 \pm 0.3$.  In contrast, values of $|C/T|$ calculated in
QCD, using a heavy-quark expansion and various assumptions
about $\Lambda_{\rm QCD}/m_b$ corrections, do not exceed 0.6
\cite{Beneke:2005vv,Li:2009wb}. The larger value of $|C/T|$
obtained in the fit \cite{Baek:2007yy} was thus considered a
possible indication for New Physics (NP)
\cite{Buras:2003dj}. (Another good fit to the data, involving
a sizable contribution from $P_{uc}$, gave a lower value
$|C/T|=0.8\pm 0.1$. However, its value of $\gamma$ was in
disagreement with the global CKM fit.)
%
\begin{table}
\caption{Branching fractions and CP asymmetries for $\btopik$ decays, as of
today and for early 2007 (in parentheses) \cite{HFAG}.
\label{tab:bkpi}}
\begin{center}
\begin{tabular}{c c c c} \hline \hline
Mode & $\b~(10^{-6})$ & $A_{CP}$ & $S_{CP}$ \\ \hline
$B^0 \to \pi^- K^+$ & $19.4 \pm 0.6$ & $-0.098^{+0.012}_{-0.011}$ & \\
& ($19.7\pm 0.6$) & ($-0.093 \pm 0.015$) & \\
$B^+ \to \pi^0 K^+$ & $12.9 \pm 0.6$ & $0.050 \pm 0.025$ & \\
& ($12.8\pm 0.6$) & ($0.047\pm 0.026$) & \\
$B^0 \to \pi^0 K^0$ &  $9.8 \pm 0.6$  & $-0.01 \pm 0.10$
 & $0.57\pm 0.17$  \\
& ($10.0\pm 0.6$) & ($-0.12\pm 0.11$) & ($0.33 \pm 0.21$) \\
$B^+ \to \pi^+ K^0$ & $23.1 \pm 1.0$  & $0.009 \pm 0.025$  & \\ 
& ($23.1 \pm 1.0$) & ($0.009\pm 0.025$) & \\
\hline \hline
\end{tabular}
\end{center}
\end{table}

Recently, an update of the fit to $\btopik$ data was
performed using the latest information as of early 2009
\cite{Baek:2009pa}.  The current data are summarized in
Table~\ref{tab:bkpi} \cite{HFAG}, compared with the data of
two years ago in parentheses. The only significant change
occurred in the central values of the two CP asymmetries
$A_{CP}$ and $S_{CP}$ in $B^0\to \pi^0K^0$, which involve the
largest experimental errors. 
The current experimental situation with respect to
the need for NP has now become less clear. Although the best
fit~\cite{Baek:2009pa}, including $P_{uc}$ as a parameter and 
using $\gamma$ as an input, is relatively poor [corresponding to
$\chi^2_{min}/d.o.f. =3.2/2~(20\%)$], it now requires
$|C/T|=0.58\pm 0.24$. This value is consistent 
with QCD calculations (within their uncertainties) and with flavor
SU(3) fits combining $B\to \pi K$ and $B\to\pi\pi$ data
\cite{Chiang:2004nm}.

A test for the CKM framework based on the asymmetry 
$S_{CP}(\pi^0K^0)$ was suggested in Ref.~\cite{Fleischer:2008wb}.
Assuming flavor SU(3) in order to obtain the quantity $|T+C|$ from
the decay rate for $B^+\to \pi^+\pi^0$~\cite{Gronau:1994bn}, the 
asymmetry $S_{CP}(\pi^0K^0)$ can be determined using as inputs  
this rate measurement and all the $B\to \pi K$ measurements except 
$S_{CP}(\pi^0K^0)$. It was noted, however, that this test is 
sensitive to input values of $\b(B^0\to \pi^0 K^0)$, 
$\b(B^0\to\pi^- K^+)$ and to SU(3) 
breaking in $\delta_{EW}$~\cite{Gronau:2008gu}.
  
The branching fractions in Table \ref{tab:bkpi} and the lifetime
ratio \cite{HFAG} $\tau_+/\tau_0 \equiv \tau(B^+)/$ \\
$\tau(B^0) = 1.073 \pm
0.008$ imply that the corresponding decay widths are
approximately in the ratio 2:1:1:2. This indicates the
dominance of the isospin-preserving ($\Delta I = 0$)
amplitude $P_{tc}$. Indeed, in the best fit of
Ref.~\cite{Baek:2009pa}, $|T|$ is about $12\%$ of $|P_{tc}|$
while other contributions are smaller. Assuming that $C$ is
suppressed relative to $T$ and can be neglected, it was shown
\cite{Gronau:1998ep} that the rate differences
\beq\label{Del-+}
\Delta(\pi^- K^+) \equiv \Gamma(\bar B^0 \to \pi^+ K^-) - 
 \Gamma(B^0 \to \pi^- K^+)
\eeq
and
\beq\label{Del0+}
2 \Delta(\pi^0 K^+) \equiv 2[\Gamma(B^- \to \pi^0 K^-) -
 \Gamma(B^+ \to \pi^0 K^+)]
\eeq
should be approximately equal.  As the CP-averaged rate for
$B^0 \to \pi^- K^+$ is about double that for $B^+ \to \pi^0
K^+$, this translates to a prediction of approximately equal
CP asymmetries,
\beq
A_{CP}(\pi^-K^+) \approx A_{CP}(\pi^0 K^+)~.
\eeq

The current measured asymmetries given in
Table~\ref{tab:bkpi} differ by more than $5\sigma$.  This has
occasionally been taken to indicate the presence of NP in
$\btopik$ \cite{Nature}.  It was noted, however
\cite{Gronau:2006xu}, that a nonzero contribution of $C$ at a
level suggested in SU(3) fits \cite{Chiang:2004nm} (and
recently in the $\btopik$ fit \cite{Baek:2009pa}) could
account for this difference. A necessary requirement is that
the strong phase difference ${\rm arg}(C/T)$ is large and
negative \cite{Gronau:2006ha}.  A large negative phase has
been obtained in flavor SU(3) fits to $\btopik$ and
$B\to\pi\pi$ data \cite{Chiang:2004nm}.  A phase of about
$-130^\circ$, obtained in the best fit in
Ref.~\cite{Baek:2009pa}, and the central value $|C/T|=0.58$
account well for the difference between the two
asymmetries. While this magnitude of $C/T$ can be accounted
for in QCD calculations, its large negative phase seems a
problem for certain QCD calculations
\cite{Bauer:2004tj,Beneke:2005vv} but not for others
\cite{Li:2009wb}.
 
A very robust sum rule was suggested a few years
ago \cite{Gronau:2005kz} combining all four $CP$ rate
asymmetries in $\btopik$:
\beq \label{eqn:sr}
\Delta(\pi^- K^+) + \Delta(\pi^+ K^0)
 \approx 2 \Delta(\pi^0 K^+) + 2 \Delta(\pi^0 K^0)~,
\eeq
where $\Delta(\pi^+ K^0)$ and $\Delta(\pi^0 K^0)$ are defined
similarly to $\Delta(\pi^- K^+)$ and $\Delta(\pi^0 K^+)$ in
Eqs.~(\ref{Del-+}) and (\ref{Del0+}). This approximate sum
rule is based on the amplitude isospin relation
[Eq.~(\ref{Iso})] and its CP conjugate, in which the
isoscalar ($\Delta I=0$) and isovector ($\Delta I=1$) parts vanish 
separately. Consider the
difference between the left-hand side of Eq.~(\ref{eqn:sr})
and its right-hand side.  The leading terms in this
difference involve interference between the dominant
isoscalar amplitude $P_{tc}$ and a linear combination of
smaller isoscalar and isovector amplitudes, each of which
vanishes because of Eq.~(\ref{Iso}). The remaining terms arise 
from the interference of smaller $\Delta I=1$
amplitudes with one another. These terms were found to vanish 
in the flavor SU(3) and heavy-quark limits \cite{Gronau:2005kz}.  A
simplified and slightly less precise version of the sum rule
[Eq.~(\ref{eqn:sr})], which assumes that the rates in Table
\ref{tab:bkpi} are in the ratio 2:1:1:2, relates the CP
asymmetries directly:
\beq \label{eqn:srACP}
A_{CP}(\pi^- K^+) + A_{CP}(\pi^+ K^0)
 \simeq A_{CP}(\pi^0 K^+) + A_{CP}(\pi^0 K^0)~.
\eeq

The sum rule in Eq.~(\ref{eqn:sr}) holds also in the presence
of a $\Delta I=0$ NP amplitude which can be absorbed in
$P_{tc}$ in the above argument.  Furthermore, since the sum
rule can only be violated by terms which are quadratic in
$\Delta I=1$ amplitudes, a substantial violation of the sum
rule requires isovector NP amplitudes which are not much
smaller than $P_{tc}$.  This will be the basis of our
following diagnostic for NP in $\btopik$ decays.

While any evidence for NP in $\btopik$ is currently weak,
suppose that NP is present in these decays. It was argued in
Ref.~\cite{DLNP} that all NP strong phases are negligible.
Given that all strong phases are equal,
there are two classes of NP amplitudes
contributing to $\btopik$, differing only in their color
structure \cite{BNPmethods}:
\bea
\ApNPqph & \equiv & \sum \bra{\pi K} {\bar s}_\alpha \Gamma_i
b_\alpha \, {\bar q}_\beta \Gamma_j q_\beta \ket{B} ~, \\
\ApNPCqph & \equiv & \sum \bra{\pi K} {\bar s}_\alpha
\Gamma_i b_\beta \, {\bar q}_\beta \Gamma_j q_\alpha \ket{B}
~,
\eea
where $\Gamma_{i,j}$ represent Lorentz structures and $q =
u,d$. (Despite the index $C$ standing for color-suppression,
the matrix elements $\ApNPCqph$ are not necessarily smaller
than $\ApNPqph$.)  Here, $\Phi_q$ and $\Phi_q^{ { C}}$ are
the NP weak phases; the strong phases are taken to be zero.
The NP amplitudes $\ApNPqph$ and $\ApNPCqph$ are equivalent  
to the amplitudes $\Delta P_q$ and $\Delta P^c_q$ defined in 
Ref.~\cite{Gronau:2005gz}.

There are three NP matrix elements which contribute to the
$\btopik$ amplitudes: $\ApNPcomb \equiv - \ApNPuph +
\ApNPdph$, $\ApNPCuph$, and $\ApNPCdph$ \cite{BNPmethods}.
(These are equivalent to the three independent NP combinations, 
$-\Delta P_u + \Delta P_d$, $\Delta P_u^c$, and $\Delta P_d^c$, contributing 
to $\btopik$ amplitudes as discussed in Ref.~\cite{Gronau:2005gz}.)
The first operator corresponds to including NP only in the
color-favored electroweak penguin amplitude: $\ApNPcomb
\equiv -\pewpnp \, e^{i \Phi_{ EW}}$. Nonzero values of
$\ApNPCuph$ and/or $\ApNPCdph$ imply the inclusion of NP in
both the gluonic and color-suppressed electroweak penguin
amplitudes, $P_{ NP} \, e^{i \Phi_{ P}}$ and $\pewcpnp \,
e^{i \Phi^{ { C}}_{ EW}}$, respectively \cite{Baek:2006ti}:
\bea
P_{ NP} \, e^{i \Phi_{ P}} & \equiv & \frac13
\ApNPCuph\ + \frac23 \ApNPCdph~, \\
\pewcpnp \, e^{i \Phi^{ { C}}_{ EW}} & \equiv &
\ApNPCuph\ - \ApNPCdph ~.
\eea

Thus, NP in $\btopik$ is of one of the above three varieties.  It
can affect the gluonic penguin amplitude ($P_{NP} \, e^{i
\Phi_P}$), the color-favored electroweak penguin amplitude
($\pewpnp \, e^{i \Phi_{EW}}$), or the color-suppressed
electroweak penguin amplitude ($\pewcpnp \, e^{i \Phi^C_{EW}}$).  
These three NP amplitudes are added to $P_{tc}, P_{EW}$ 
and $P^C_{EW}$, respectively.
In the first case, the NP is $\Delta I=0$ and,
as mentioned, will not affect the sum rule of
Eq.~(\ref{eqn:sr}).  In the other two cases, the NP breaks
isospin by one unit, so that the sum rule will be violated.
Ref.~\cite{Baek:2009pa} performed fits with each of the three
NP operators.  In the SM-like fit it was found that it is not possible to
constrain NP in the $\Delta I=0$ gluonic penguin 
($P_{ NP} \,e^{i \Phi_{ P}}$).  This is not surprising -- it is the same
in $B \to \pi\pi$ decays \cite{pipiI=0}.  However, the two non-SM-like
fits did indeed involve substantial contributions from
NP: $|\pewpnp/P_{tc}| = 0.4$, $|\pewcpnp/P_{tc}| = 0.3$. They
had lower values of $\chi^2/d.o.f.$ (0.4/2 and 2.5/2)
than the SM-like fit (3.6/2).  One potentially troublesome point
is that they had very large values of $|C/T|$ ($6 \pm 11$ and
$4.9 \pm 3.8$).  However, the errors are also quite large, so
this might not be a real problem.  The key point here is that
these two NP operators are of the type $\Delta I=1$, and so
can violate the sum rules of Eqs.~(\ref{eqn:sr}) and
(\ref{eqn:srACP}).

Using the CP asymmetries [all except for $A_{CP}(B^0 \to
\pi^0 K^0)$] and branching ratios in Table~\ref{tab:bkpi},
and translating ratios of branching ratios into ratios of
rates using the lifetime ratio $\tau_+/\tau_0 = 1.073
\pm 0.008$ \cite{HFAG}, the sum rule of Eq.~(\ref{eqn:sr})
predicts
\beq
A_{CP}(B^0 \to \pi^0 K^0) = - 0.149 \pm 0.044~,
\eeq
more than $3 \sigma$ away from zero.  The $B^0 \to \pi^0 K^0$
mode is thus predicted to exhibit the largest CP asymmetry of
any of the four $B \to \pi K$ modes.

If one uses the simplified version of the sum rule
[Eq.~(\ref{eqn:srACP})], which assumes the rates in Table
\ref{tab:bkpi} are in the ratio 2:1:1:2, the CP asymmetries
are related directly:
\beq \label{eqn:sr1}
A_{CP}(\pi^0 K^0) = A_{CP}(\pi^- K^+) + A_{CP}(\pi^+ K^0) - A_{CP}(\pi^0
K^+) = -0.139 \pm 0.037~.
\eeq
Indeed, the SM-like fit in Ref.\ \cite{Baek:2009pa} finds
$A_{CP}(\pi^0 K^0) = -0.12$.  (We restrict our attention to
those fits which include the constraints on CKM phases
\cite{HFAG} $\beta = (21.66^{+0.95}_{-0.85})^\circ$ and
$\gamma = (66.8^{+5.4}_{-3.8})^\circ$.)

On the other hand, in the other two fits of Ref.\ \cite{Baek:2009pa},
$\Delta I=1$ NP amplitudes are involved, which are much more significant
than those in the SM, and very large values of $|C|$ are obtained.  This
results in a crucial difference with the SM fit: both of
these fits predict $A_{CP}(\pi^0 K^0) = -0.03$.  We thus see
that it is possible to detect the presence of NP with a
precise measurement of $A_{CP}(\pi^0 K^0)$.

Another sum rule involving decay rates instead of CP asymmetries
is, in principle, sensitive to NP in $\Delta I = 1$ amplitudes
\cite{Gronau:1998ep,Lipkin:1998ie}.  In terms of branching ratios
$\b(B \to f) \equiv \b(f)$ it is expressed as
\beq \label{eqn:GRL}
2\b(\pi^0 K^+) + 2(\tau_+/\tau_0)\b(\pi^0 K^0) =
 (\tau_+/\tau_0)\b(\pi^- K^+) + \b(\pi^+ K^0)~. 
\eeq
The difference between the left- and right-hand sides is quadratic
in $\Delta I = 1$ amplitudes, expected to be small in the Standard
Model.  A discussion of the terms violating this sum rule can be 
found in Ref.\ \cite{Gronau:2005gz}, which also discusses NP in
$B \to \pi K$ decays.

The experimental branching ratios in Table \ref{tab:bkpi}  satisfy the sum
rule at the $1.4 \sigma$ level, with the (left, right)-hand sides giving
$(46.8\pm1.8,~43.9\pm1.2)$ in units of $10^{-6}$.  The three fits of
Ref.\ \cite{Baek:2009pa} satisfy the sum rule as well or better, as shown
in Table \ref{tab:comp}.  A large $P_{EW,NP}$ amplitude in Fit 2 is nearly
cancelled by a large $C$ contribution in a delicate way so as to preserve the
sum rule.

\begin{table}
\caption{Comparison of fits of Ref.\ \cite{Baek:2009pa} with and without
constraint $|C/T|=0.5$. Prediction for (a) $A_{CP}(\pi^0 K^0)$, (b) Eq.~(\ref{eqn:GRL}),
l.h.s; (c) Eq.~(\ref{eqn:GRL}), r.h.s.
\label{tab:comp}}
\begin{center}
\begin{tabular}{c c c c c c c c c} \hline \hline
Fit & \multicolumn{4}{c}{With $|C/T|=0.5$} & \multicolumn{4}{c}{Without
 $|C/T|=0.5$} \\
    & $\chi^2/d.o.f.$ & (a) & (b) & (c) & $\chi^2/d.o.f.$ & (a) & (b) & (c) \\ \hline
 1 & 3.7/3 & $-0.11$ & 44.9 & 44.8 & 3.6/2 & $-0.12$ & 44.9 & 44.8 \\
 2 & 3.0/3 & $-0.12$ & 47.1 & 43.2 & 0.4/2 & $-0.03$ & 47.0 & 43.8 \\
 3 & 3.8/3 & $-0.12$ & 44.9 & 44.8 & 2.5/2 & $-0.03$ & 45.3 & 44.6 \\ \hline \hline
\end{tabular}
\end{center}
\end{table}

One may ask if the large value of $|C/T|$ in Fits 2 and 3 is obligatory.
Performing the same fits but constraining $|C/T|=0.5$, one finds results
summarized in Table \ref{tab:comp}.  The number of degrees of freedom
with the constraint $|C/T|=0.5$ is 3, while without the constraint it is 2.
It appears that a large value of $|C|$, along with a sizable NP amplitude,
is responsible for moving the
predicted $A_{CP}(\pi^0 K^0)$ away from its Standard Model value of $-0.15$. 

The difference between the violation of the rate sum rule
[Eq.~(\ref{eqn:GRL})] and the asymmetry sum rule
[Eq.~(\ref{eqn:sr})] in the presence of NP can be seen as
follows.  Define $P_1 \equiv P_{EW,NP}$ and $P_2 \equiv
P^C_{EW,NP}$ and assume they are large, neglecting the
violation of the two sum rules by small SM contributions
(which is a good approximation).

The terms violating the rate sum rule are
\beq
2[|P_1|^2 + |P_1||P_2| \cos(\delta_1 - \delta_2) \cos(\phi_1 - \phi_2)]~,
\eeq
where $\delta_1,\delta_2$ and $\phi_1,\phi_2$ are the strong
and weak phases of $P_1$ and $P_2$.  This leads to a
violation of the rate sum rule [Eq.~(\ref{eqn:GRL})] in the
case $P_1 \ne 0$, $P_2 = 0$ (Fit 2 of Ref.\
\cite{Baek:2009pa}) but to no violation in the case $P_1 =
0$, $P_2 \ne 0$ (Fit 3 of Ref.\ \cite{Baek:2009pa}).

On the other hand, the term violating the asymmetry sum rule is
\beq
2|P_1||P_2| \sin(\delta_1 - \delta_2) \sin(\phi_1 - \phi_2)~.
\eeq
In Ref.\ \cite{Baek:2009pa} it was assumed that $\delta_1 = \delta_2$,
so the asymmetry sum rule was not violated by this term.  Also, the
cases $P_1 = 0$, $P_2 \ne 0$ and $P_1 \ne 0$, $P_2 = 0$ were considered
separately.  In these cases the sum rule would hold even when $\delta_1 \ne
\delta_2$.

The reason for the violation of the asymmetry sum rule in the two fits
(Fit 2 and Fit 3) of Ref.\ \cite{Baek:2009pa} without a constraint on
$|C/T|$ (seventh column, second and third rows of Table \ref{tab:comp}),
giving $A_{CP} = -0.03$ instead of $-0.12$, is that $C$ was also large
in both fits, so the sum rule was violated by contributions
\bea
4|P_1||C|\,\sin(\delta_1 - \delta_C)\, \sin(\phi_1 -
\gamma)~~~~({\rm Fit~2})~,\\
2|P_2||C|\,\sin(\delta_2 - \delta_C)\,\sin(\phi_2 -
\gamma)~~~~({\rm Fit~3})~.
\eea

To sum up, the asymmetry sum rule [Eq.~(\ref{eqn:sr})] can be
violated significantly (when assuming $|C/T|$ is not very
large) only by taking $P_1 \ne 0$, $P_2 \ne 0$, $\delta_1 \ne
\delta_2$, and $\phi_1 \ne \phi_2$.  In other words,
violation of the sum rule requires both $P_{EW,NP}$ and
$P^C_{EW,NP}$ to be present and both strong- and weak-phase
differences of these amplitudes to be non-negligible.  One
may imagine a situation in which these circumstances hold
leading to an asymmetry $A_{CP}=-0.03$ in $B^0\to\pi^0 K^0$
as discussed above.

Although we recognize the difficulty of the measurement, we
thus urge a reduction of the present experimental error on
$A_{CP}(\pi^0 K^0)$, whose world average is $-0.01 \pm 0.10$.
The BaBar and Belle contributions to this average are
compared in Table \ref{tab:p0k0}. The error must be reduced
by at least a factor of three in order to determine whether
new physics is generating a large $\Delta I = 1$ transition
amplitude.  An experiment collecting of the order of
$10^{10}$ $B \bar B$ pairs ought to be able to make the
necessary distinction.

\begin{table}
\caption{Comparison of BaBar \cite{2008se} and Belle \cite{Fujikawa:2008pk}
values of $A_{CP}(B^0 \to \pi^0 K^0)$.
\label{tab:p0k0}}
\begin{center}
\begin{tabular}{c c c} \hline \hline
Source & $N(B \bar B)$ (M) & $A_{CP}$ \\ \hline
BaBar \cite{2008se} & 467 & $-0.13 \pm 0.13 \pm 0.03$ \\ 
Belle \cite{Fujikawa:2008pk} & 657 & $0.14 \pm 0.13 \pm 0.06$ \\
Average \cite{HFAG}  & 1124 & $-0.01 \pm 0.10$ \\ \hline \hline
\end{tabular}
\end{center}
\end{table}

\bigskip
{\it Acknowledgments:} This work was financially supported in
part by the Korea Research Foundation Grant funded by the
Korean Government (MOEHRD) No. KRF-2007-359-C00009 (SB), by
the National Science Council of Taiwan, R.O.C. under Grant
No.~NSC 97-2112-M-008-002-MY3 (CC), by NSERC of Canada (DL),
and by the United States Department of Energy through Grant
No.\ DE FG02 90ER40560 (JR).

\end{document}